\documentclass{article}
\usepackage{graphicx}  
\usepackage{amsmath}   
\usepackage[compress]{cite}
\usepackage{amssymb}   
\usepackage{bm} 
\usepackage{dcolumn}
\usepackage{color}
\usepackage{mathrsfs}
\usepackage{amsfonts}
\usepackage{varioref}
\RequirePackage[colorlinks,citecolor=blue,urlcolor=magenta,linkcolor=blue]{hyperref}
\def\EH{Einstein-Hilbert }
\def\LL{Lanczos-Lovelock }
\labelformat{section}{Section #1} 
\labelformat{subsection}{Section #1} 
\labelformat{subsubsection}{Section #1}
\labelformat{subsubsubsection}{Section #1}
\labelformat{equation}{Eq.~(#1)} 
\labelformat{figure}{Fig.~#1} 
\labelformat{subfigure}{Fig.~\thefigure#1} 
\labelformat{table}{Tab.~#1} 
\labelformat{appendix}{Appendix #1}
\title{Solving higher curvature gravity theories}
\author{Sumanta Chakraborty
\footnote{sumanta@iucaa.in;~~sumantac.physics@gmail.com}\\
{\small{IUCAA, Post Bag 4, Ganeshkhind, Pune University Campus, Pune 411 007, India}}\\
and\\
Soumitra SenGupta
\footnote{tpssg@iacs.res.in}\\
{\small{Theoretical Physics Department, Indian Association for the Cultivation of Science, Kolkata 700032, India}}}
\begin{document}
  
\maketitle
\begin{abstract}
Solving field equations in the context of higher curvature gravity theories is a formidable task. However in many situations, e.g., in the context of $f(R)$ theories the higher curvature gravity action can be written as Einstein-Hilbert action plus a scalar field action. We show that not only the action but the field equations derived from the action are also equivalent provided the spacetime is regular. We also demonstrate that such equivalence continues to hold even when gravitational field equations are projected on a lower dimensional hypersurface. We have further depicted explicit examples in which the solutions for Einstein-Hilbert and a scalar field system lead to solutions of the equivalent higher curvature theory. The same, but on the lower dimensional hypersurface, has been illustrated in the reverse order as well. We conclude with a brief discussion on this technique of solving higher curvature field equations.
\end{abstract}
\section{Introduction}\label{Intro_fR}

The energy scales in particle physics are arranged in a hierarchical manner. While the scale of Weak interaction corresponds to $E\sim 10^{3}~\textrm{GeV}$, the Strong interaction at a scale of $E\sim 10^{16}~\textrm{GeV}$ exceeds the Weak scale by a factor of $10^{13}$. This large difference leads to a fine tuning problem in the scheme of renormalization --- known as the \textit{gauge hierarchy problem}. This fine tuning is absolutely necessary to renormalize the mass of Higg's Boson, which recently have been detected with a mass of $127~\textrm{GeV}$. At face value, this fine tuning is what nature prefers and the question is \textit{why?} Hence it is natural to ask, is there a more fundamental principle from which this fine tuning would appear naturally. 

There have been a large amount of work to address the hierarchy problem, a few candidates have emerged out of it --- supersymmetry, technicolor and extra dimensions. In this work we will be concerned solely about the third alternative, i.e., we will assume the actual spacetime has more than three spatial dimensions (commonly referred to as the \emph{bulk}), while the spacetime we live in is a four-dimensional hypersurface in the bulk (commonly referred to as the \emph{brane}). The two immediate observable consequences are --- change of $1/r^{2}$ gravitational force law at the length scale of extra dimensions, existence of massive graviton through Kaluza-Klein tower \cite{ArkaniHamed:1998rs,Antoniadis:1998ig,Antoniadis:1990ew,Rubakov:1983bb,Horava:1996ma,Kaloper:1999sm,
Cvetic:1996vr,Cohen:1999ia,Maia:2001gq}.

To probe these extra dimensions one needs to have high enough energy or high enough curvature, such that the relevant energy scale of the problem becomes close to the Planck scale. General Relativity, described by the \EH action is considered to be an effective theory of gravity, valid much below the fundamental Planck scale \cite{Buchbinder1992}. Once energies approach the Planck scale, one not only expects to observe deviations from the \EH action but also signatures of the extra dimensions. This is particularly relevant, since future colliders will probe higher and higher energies such that aspects beyond general relativity should become apparent. Since the ultraviolet behaviour of the true gravity theory is yet unknown one hopes that in these high energy/high curvature regimes deviations from standard model or deviations from Einstein gravity may appear through existence of extra dimensions. To capture some of the aspects of ``quantum gravity'' one is tempted to consider how the presence 
of higher curvature (and higher derivative) invariants in the higher dimensional gravitational action modifies the well-known results \cite{Rizzo:2006wf,Konoplya:2008ix,Brown:2007dd,Rizzo:2006sb,Cai:2006pq}.

The higher derivative terms that one can add to the \EH action are not unique. However many of these terms can lead to a linear instability, called Ostrogradski instability, leading to appearance of ghost fields and hence will not be considered in this work. Among the higher curvature theories, the \LL gravity and the $f(R)$ gravity are of much importance. The \LL gravity is special in the sense that the field equations derived from the \LL action contain only second order derivatives of the metric and have natural thermodynamic interpretation \cite{Chakraborty:2015wma,Chakraborty:2015hna,Chakraborty:2015kva,Chakraborty:2014rga,
Chakraborty:2014joa}. On the other hand, $f(R)$ gravity was first introduced to explain both early and late time exponential expansion of the universe without invoking additional matter components, e.g., dark energy \cite{Nojiri:2001ae,Nojiri:2010wj,Sotiriou:2008rp,DeFelice:2010aj,Paliathanasis:2015aos,Basilakos:2011rx,Zhong:2010ae}. But only addressing cosmological observations do \emph{not} lead to a viable model, for that the $f(R)$ theories should pass the local gravity tests --- perihelion precession of Mercury and bending angle of light as well. It turns out that Solar System experiments do not exclude the viability of $f(R)$ theories at scales shorter than the cosmological ones, but provide constraints on $f''(R)$ and hence constraints on the parameters of the model. Thus it is can be affirmed that extended gravity theories cannot be ruled out, definitively, using Solar System experiments \cite{Capozziello:2005bu,Capozziello:2007ms,Sotiriou:2005xe,Capozziello:2006jj}. 

It is also well known that $f(R)$ gravity theories can be related to scalar-tensor theories by a conformal transformation at the action level \cite{Barrow:1988xh,Capozziello:1996xg,Nojiri:2010wj,Sotiriou:2008rp,DeFelice:2010aj,Anand:2014vqa,Bahamonde:2016wmz,
Parry:2005eb,Catena:2006bd,Chiba:2013mha}. Thus it is important to consider the following situation --- obtaining field equations from the scalar-tensor representation and from $f(R)$ gravity representation. Since the two actions are related by a conformal transformation, the field equations should also be equivalent. However the situation is \emph{not} trivial, since the metric in scalar-tensor representation depends on the conformal factor, its variation can potentially lead to various additional terms, which must cancel other terms \emph{exactly} in order to arrive at the equivalence. If the equivalence exist, we can use it to solve field equations for scalar-tensor theory and obtain the solution corresponding to $f(R)$ action and vice-versa. This would be advantageous, since in \emph{general} solving the field equations for $f(R)$ gravity, where $R$ is not a constant, is difficult\footnote{Note that in cosmological context one can solve the field equations for $f(R)$ gravity by a trick, 
known as the \emph{reconstruction method} \cite{Nojiri:2006gh,Capozziello:2006dj}. 
We will have occasion to comment on this method when we compare our technique introduced in this 
work with the \emph{reconstruction scheme}.} \cite{Nojiri:2010wj,Chakraborty:2014zya,Bhattacharya:2016lup,Capozziello:2011et,Barausse:2007pn,
Capozziello:2009nq,Aghamohammadi:2010yq,Capozziello:2011wg}. While the corresponding scalar-tensor solution could in principle be much simpler. The same should work on the brane hypersurface as well. The effective field equations on the brane derived through Gauss-Codazzi formalism in the $f(R)$ representation \cite{Chakraborty:2014xla,Dadhich:2000am,Shiromizu:1999wj,Harko:2004ui,Chakraborty:2015bja,Chakraborty:2015taq} should be equivalent to the same but derived from the scalar-tensor representation. The non-triviality of this result originates from the quadratic combination of energy momentum tensor and extrinsic curvature appearing in effective field equations.

As an aside, we should mention that the conformal transformation is well motivated \emph{only} when the spacetime does not have singularities. Such singularity free spacetimes have been obtained earlier in the context of cosmology with moduli dependent loop corrections of the gravitational part of superstring effective action with orbifold compactifications \cite{Antoniadis:1993jc}. However to obtain a singularity free description it was necessary that the stress energy tensor associated with the modulus should violate the strong energy condition. In circumstances, where the energy conditions are obeyed, one obtains singular solutions in general. For singular spacetimes, viz., cosmological spacetimes near the Big Rip or Big Crunch the transformation can break down. In those contexts it can exhibit peculiar behavior, e.g., the Big Rip singularity, which may appear in some versions of $f(R)$ gravity can either map itself to infinite past or future, or can be replaced by a Big Crunch singularity 
\cite{Briscese:2006xu,Bahamonde:2016wmz}. Another point requires clarification at this stage, this has to do with physical non-equivalence of the two frames. All the comments phrased above has to do with mathematical equivalence, but the physical solutions can be very different \cite{Briscese:2006xu}. This is evident, since the conformal factor can change the complete structure of the spacetime. This fact was pointed out earlier in \cite{Dabrowski:2008kx} by showing that through conformal transformation one can create matter and as a result, one frame is empty while another has matter, clearly they are physically non-equivalent. This should not come as a surprise, since Schwarzschild metric under conformal transformation no longer satisfies Einstein's equations. Further in the cosmological context for $f(R)$ gravity model it was explicitly demonstrated \cite{Briscese:2006xu,Capozziello:2010sc} that neither the Hubble parameter nor the deceleration parameter matches in Jordan and Einstein frame, showing the 
physical non-equivalence. In view of the above, the phrase ``equivalence'' in the following sections should be understood in a mathematical sense, \emph{not} in a physical sense. Further, we will contend ourselves, with \emph{only those spacetimes (or regions of spacetimes) which are regular}, such that the conformal transformation between the two frames is well defined throughout the region of interest.

The paper is organized as follows: In \ref{Sec_01} we present a brief review of the equivalence between the $f(R)$ gravity and scalar-tensor theory in five dimensions and hence the equivalence between the bulk field equations as well. \ref{Sec_02} is devoted to show the equivalence between the effective field equations on the brane. The application of the bulk equivalence is presented in \ref{Sec_03}. There we have started from scalar-tensor theory and have solved the bulk equations, from which the solution in $f(R)$ representation is obtained. In \ref{Sec_04} we consider brane spacetime where the solution in scalar-tensor representation starting from $f(R)$ representation is derived as another explicit example. We conclude with a brief discussion on this technique. All the relevant calculations are presented in \ref{GBAPP_01}. 

We have set the fundamental constants, $c$ and $\hbar$ to unity and shall work with mostly positive signature of the metric. The Latin indices, $a,b,\ldots$ runs over the full spacetime indices, while Greek indices, $\mu,\nu,\ldots$ stands for four dimensional spacetime. 
\section{Equivalence of gravitational field equations in the bulk}\label{Sec_01}

The starting point for any fundamental theory corresponds to correct identification of the dynamical variable and the associated field equations. An useful trick to obtain the field equations is to introduce an action principle, extremizing which one can obtain the field equations. Along similar lines, in gravitational theories as well one considers the metric $g_{ab}$ as dynamical. Given an appropriate action, when arbitrary variations of $g_{ab}$ are considered, the gravitational action reaches extremum value only if $g_{ab}$ satisfies the gravitational field equations. For example, Einstein's equations follow from variation of the Einstein-Hilbert action, which is the Ricci scalar. At high enough curvature (or energy) the \EH action is most likely to be supplemented by higher curvature corrections. Among many such viable modifications, $f(R)$ theories are of particular interest. The action for $f(R)$ gravity model (also known as the Jordan frame) in five spacetime dimensions reads,
\begin{align}\label{J_Action}
S_{J}\equiv \int d^{5}x\sqrt{-g}\left[\frac{f(R)}{2\kappa _{5}^{2}}+\mathcal{L}_{m}\right]=\int d^{5}x\sqrt{-g}\left[\left(\frac{f'R}{2\kappa _{5}^{2}}-\frac{f'R-f}{2\kappa _{5}^{2}}\right)+\mathcal{L}_{m}\right]~,
\end{align}
where $\kappa _{5}$ is the five dimensional gravitational constant, $\mathcal{L}_{m}$ is the matter Lagrangian and $f'$ stands for $df/dR$. Conformal transformation of the Jordan frame metric $g_{ab}$ results in: $\mathbf{g}_{ab}=\Omega ^{2}g_{ab}$, where $\Omega$ is the conformal factor. The resulting action, written in terms of $\mathbf{g}_{ab}$ can be obtained using the transformation properties of curvature tensor yielding (as presented in \ref{GBAPP_01}),
\begin{align}\label{E_Action_01}
S&=\int d^{5}x\sqrt{-\mathbf{g}}\left[\Omega ^{-5}\left\lbrace \frac{f'}{2\kappa _{5}^{2}}\left(\Omega ^{2}\mathbf{R}+8\Omega ^{2}\boldsymbol{\square}\ln \Omega -12\mathbf{g}^{ab}\nabla _{a}\Omega \nabla _{b}\Omega\right)-U(f)\right\rbrace +\boldsymbol{\mathcal{L}_{m}}\right]~.
\end{align}
Here $U(f)$ stands for $(f'R-f)/2\kappa _{5}^{2}$ and $\boldsymbol{\mathcal{L}_{m}}$ is the matter Lagrangian in the conformally transformed action. Note that, in our convention, all the metric dependent quantities originating from conformal transformation of the Jordan frame are boldfaced. We will follow this convention throughout this work. Under the following identifications
\begin{align}\label{conditions_01}
\Omega ^{3}=f';\qquad \kappa _{5}\phi =\frac{2}{\sqrt{3}}\ln f'=2\sqrt{3}\ln \Omega~,
\end{align}
the action presented in \ref{E_Action_01} reduces to the following form (known as the Einstein frame action), 
\begin{align}\label{E_Action_02}
S_{E}=\int d^{5}x\sqrt{-\mathbf{g}}\left(\frac{\mathbf{R}}{2\kappa _{5}^{2}}-\frac{1}{2}\mathbf{g}^{ab}\nabla _{a}\phi \nabla _{b}\phi-V(\phi)+\boldsymbol{\mathcal{L}_{m}}\right)+\int d^{4}x \sqrt{h}\frac{4}{\kappa _{5}^{2}}n^{c}\nabla _{c}\ln \Omega~.
\end{align}
The Einstein frame action can be divided into a bulk term and a surface term, as evident from \ref{E_Action_02}. If we are only interested in variation of the action and derivation of the field equations, the boundary term can be safely ignored. However, while concentrating on brane dynamics, which is a boundary effect, the surface term will be important. Further, the potential term $V(\phi)$ appearing in \ref{E_Action_02} corresponds to,
\begin{align}\label{conditions_02}
V(\phi)=\frac{f'(\phi)R(\phi)-f(\phi)}{2\kappa _{5}^{2}f'(\phi)^{5/3}}~,
\end{align}
which can be obtained by solving \ref{conditions_01}. Extremizing the Einstein frame action presented in \ref{E_Action_02} with respect to arbitrary variations of the metric $\mathbf{g}^{ab}$ we readily obtain the field equations in the Einstein frame,
\begin{align}\label{E_bulk}
\mathbf{G}_{ab}\equiv \mathbf{R}_{ab}-\frac{1}{2}\mathbf{R}\mathbf{g}_{ab}=\kappa _{5}^{2}\left[\nabla _{a}\phi \nabla _{b}\phi-\mathbf{g}_{ab}\left(\frac{1}{2}\mathbf{g}^{cd}\nabla _{c}\phi \nabla _{d}\phi+V(\phi)\right)\right]+\kappa _{5}^{2}\mathbf{T}_{ab}~.
\end{align}
The energy momentum tensor $\mathbf{T}_{ab}$ is obtained from the matter action $\boldsymbol{\mathcal{L}_{m}}$ by variation of the Einstein frame metric $\mathbf{g}_{ab}$. Using the conformal transformation, the energy momentum tensor $\mathbf{T}_{ab}$ in the Einstein frame can be related to the energy momentum tensor $T_{ab}$ in the Jordan frame as,
\begin{align}
\mathbf{T}_{ab}&\equiv-\frac{2}{\sqrt{-\mathbf{g}}}\frac{\delta \left(\sqrt{-\mathbf{g}}\boldsymbol{\mathcal{L}_{m}}\right)}{\delta \mathbf{g}^{ab}}
\nonumber
\\
&=-\frac{2}{\sqrt{-g}\Omega ^{5}}\frac{\delta \left(\sqrt{-g}\mathcal{L}_{m}\right)}{\Omega ^{-2}\delta g^{ab}}
=\frac{1}{f'}T_{ab}~.
\end{align}
Since the Jordan frame action $S_{J}$ is derivable from Einstein frame action $S_{E}$ through conformal transformation, the bulk equations derived from them should coincide. The above statement, though physically well motivated is by no means trivial. This has to do with the fact that while deriving the Einstein frame equations one should vary the Einstein frame metric $\mathbf{g}_{ab}$. This in turn leads to arbitrary variation of the Jordan frame metric $g_{ab}$ and the conformal factor $\Omega$. Since the conformal factor can be written in terms of the curvature tensor due to the identification in \ref{conditions_01}, it can lead to various additional correction terms. It is not clear a priori, how these terms combine and yield correct field equations in the Jordan frame. Since there exist no explicit derivation of the same, it is worthwhile to explicitly demonstrate the equivalence. 

In order to prove the same we will start from \ref{E_bulk} and shall try to write every curvature tensor components in terms of the Jordan frame metric $g_{ab}$. This can be done using transformation properties of curvature tensors between the two frames related by conformal transformation, leading to,
\begin{align}
\mathbf{R}_{ab}=R_{ab}-\frac{\nabla _{a}\nabla _{b}f'}{f'}+\frac{4}{3}\frac{\nabla _{a}f'\nabla _{b}f'}{f'^{2}}-g_{ab}\frac{\square f'}{3f'}~,
\end{align}
and
\begin{align}
\mathbf{g}_{ab}\mathbf{R}=g_{ab}R-\frac{8}{3}g_{ab}\frac{\square f'}{f'}+\frac{4}{3}g_{ab}\frac{g^{cd}\nabla _{c}f'\nabla _{d}f'}{f'^{2}}~.
\end{align}
Having expressed both the Ricci tensor and Ricci scalar in the Einstein frame in terms of the conformal factor and corresponding curvature components in the Jordan frame, the Einstein tensor in the Einstein frame can be expressed as, 
\begin{align}
\mathbf{G}_{ab}=G_{ab}-\frac{\nabla _{a}\nabla _{b}f'}{f'}+\frac{4}{3}\frac{\nabla _{a}f'\nabla _{b}f'}{f'^{2}}-\frac{2}{3}g_{ab}\frac{g^{cd}\nabla _{c}f'\nabla _{d}f'}{f'^{2}}+g_{ab}\frac{\square f'}{f'}~.
\end{align}
Further the contribution from scalar field present in the right hand side of \ref{E_bulk} can be written in terms of $f(R)$ and its various derivatives as,
\begin{align}\label{Tab_phi}
\nabla _{a}\phi \nabla _{b}\phi &-\mathbf{g}_{ab}\left(\frac{1}{2}\mathbf{g}^{cd}\nabla _{c}\phi \nabla _{d}\phi+V(\phi)\right)
=\frac{1}{\kappa _{5}^{2}}\left[\frac{4}{3}\frac{\nabla _{a}f'\nabla _{b}f'}{f'^{2}}-\frac{2}{3}g_{ab}\frac{g^{cd}\nabla _{c}f'\nabla _{d}f'}{f'^{2}}-g_{ab}\frac{f'R-f}{2f'}\right]~.
\end{align}
Using these relations between Einstein frame and Jordan frame, the field equations in Einstein frame presented in \ref{E_bulk} can be written as,
\begin{align}
f'G_{ab}+\frac{f'R-f}{2}g_{ab}-\nabla _{a}\nabla _{b}f'+g_{ab}\square f'=\kappa _{5}^{2}T_{ab}~,
\end{align}
which is precisely the field equations one would have obtained by extremizing \ref{J_Action} for arbitrary variation of the Jordan frame metric $g_{ab}$. Hence follows the equivalence. As a consequence if one can solve for $\mathbf{g}_{ab}$ starting from the field equations in the Einstein frame, the solution in Jordan frame can be obtained through a conformal transformation and vice versa. We should emphasize that the above statement though mathematically correct, practically might require suitable approximations for inverting various functional relations connecting the two frames. We will provide detailed comments on this aspect later on, while providing concrete examples.

Even though we have used metric formalism to arrive at the equivalence, one can also use another method known as \emph{Palatini method}. For discussions on the same we refer our reader to \cite{Olmo:2005hc,Iglesias:2007nv,Barausse:2007ys,Sotiriou:2008rp,Olmo:2011uz}.
\section{Equivalence of effective field equations on the brane}\label{Sec_02}

In the previous section, we have shown the equivalence between the bulk field equations derived from the Einstein and the Jordan frame. However from the perspective of brane world, governed by effective field equations derived from the bulk action, the equivalence of the effective field equations are more important. The effective field equations involve various quadratic combinations of the extrinsic curvature and the matter energy momentum tensor. Thus all the additional terms with their appropriate factors present in the Einstein frame must cancel each other such that effective field equations in the Jordan frame is obtained. In this connection we would like to highlight that in most of the works related to $f(R)$ gravity the surface term in the Einstein frame is ignored, however in order to prove the equivalence on the brane this term is absolutely necessary. Hence the equivalence of effective field equations too, is a non-trivial statement. In this section we will explicitly demonstrate the same.

The bulk field equations in the Einstein frame involves energy momentum tensor of the scalar field along with any other matter fields which may be present in the bulk. The bulk energy momentum tensor $T_{ab}^{\rm bulk}$ (the trace is denoted by $T^{\rm bulk}$) will induce an effective brane energy momentum tensor $T_{ab}^{\rm brane}$ as,
\begin{align}\label{bulk_brane}
\kappa _{4}^{2}T_{\mu \nu}^{\rm brane}=\frac{2}{3}\kappa _{5}^{2}\left[T_{ab}^{\rm bulk}e^{a}_{\mu}e^{b}_{\nu}+h_{\mu \nu}\left(T_{ab}^{\rm bulk}n^{a}n^{b}-\frac{1}{4}T^{\rm bulk}\right)\right]~.
\end{align}
Here $\kappa _{5}$ is the five-dimensional (i.e., bulk) gravitational constant while $\kappa _{4}$ is the four-dimensional (i.e., brane) gravitational constant. Moreover, the object, $e^{a}_{\mu}$ stands for $\partial x^{a}/\partial y^{\mu}$, where $y^{\mu}$ corresponds to the brane coordinates and $x^{a}$ are the bulk coordinates. The normal to the brane hypersurface being $n_{a}$ such that the induced metric on the brane hypersurface becomes $h_{\mu \nu}=e^{a}_{\mu}e^{b}_{\nu}(g_{ab}-n_{a}n_{b})$ \cite{Poisson}.

To obtain the effective field equations in the Jordan frame, we need to express the scalar field in terms of $f(R)$ and its derivatives. Thus, the bulk energy momentum tensor $\mathbf{T}_{ab}^{(\phi)}$ for the scalar field reads,
\begin{align}
\kappa _{5}^{2}\mathbf{T}_{ab}^{(\phi)}=\frac{4}{3}\nabla _{a}\ln f'\nabla _{b}\ln f'-\frac{2}{3}g_{ab}\left(g^{cd}\nabla _{c}\ln f'\nabla _{d}\ln f'\right)-g_{ab}\frac{f'R-f}{2f'}~.
\end{align}
Using which the following results can be obtained
\begin{align}
\kappa _{5}^{2}\mathbf{T}_{ab}^{(\phi)}\mathbf{n}^{a}\mathbf{n}^{b}&=\frac{1}{\Omega ^{2}}\Big[\frac{4}{3}\left(n^{a}\nabla _{a}\ln f'\right)^{2}-\frac{2}{3}\nabla _{a}\ln f'\nabla ^{a}\ln f'-\frac{f'R-f}{2f'}\Big]~,
\\
\kappa _{5}^{2}\mathbf{T}^{(\phi)}&=\frac{1}{\Omega ^{2}}\Big[\frac{4}{3}\nabla _{a}\ln f'\nabla ^{a}\ln f'-\frac{10}{3}\nabla _{a}\ln f'\nabla ^{a}\ln f'-5\frac{f'R-f}{2f'}\Big]~.
\end{align}
From these expressions of the bulk energy momentum tensor of the scalar field, the brane energy momentum tensor, after some simplifications (using \ref{bulk_brane} in particular), leads to, 
\begin{align}
\kappa _{4}^{2}\mathbf{T}^{(\phi)\textrm{brane}}_{\mu \nu}&=\frac{8}{9}\nabla _{\mu}\ln f'\nabla _{\nu}\ln f'-\frac{5}{9}h_{\mu \nu}\nabla _{a}\ln f'\nabla ^{a}\ln f'
\nonumber\
\\
&-\frac{3}{4}h_{\mu \nu}\frac{f'R-f}{3f'}+\frac{8}{9}h_{\mu \nu}\left(n^{a}\nabla _{a}\ln f'\right)^{2}~.
\end{align}
Let us now work through the last bit of this analysis regarding Einstein tensor. Using the transformation properties of Riemann tensor and Ricci scalar (see for example, \ref{GBAPP_01}) we immediately obtain the following result for the induced Einstein tensor,
\begin{align}
\mathbf{G}_{\mu \nu}=G_{\mu \nu}-2\frac{\nabla _{\mu}\nabla _{\nu}\Omega}{\Omega}+2h_{\mu \nu}\frac{h^{\alpha \beta}\nabla _{\alpha}\nabla _{\beta}\Omega}{\Omega}+4\frac{\nabla _{\mu}\Omega \nabla _{\nu}\Omega}{\Omega ^{2}}-h_{\mu \nu}\frac{h^{\alpha \beta}\nabla _{\alpha}\Omega \nabla _{\beta}\Omega}{\Omega ^{2}}~.
\end{align}
In this particular case, the conformal factor $\Omega$ is related to $df/dR$ through the relation $\Omega=f'^{1/3}$. Using this relation, after some straightforward manipulation and simplification we arrive at,
\begin{align}
\mathbf{G}_{\mu \nu}=G_{\mu \nu}-\frac{2}{3}\frac{\nabla _{\mu}\nabla _{\nu}f'}{f'}+\frac{8}{9}\frac{\nabla _{\mu}f'\nabla _{\nu}f'}{f'^{2}}
-\frac{5}{9}h_{\mu \nu}h^{\alpha \beta}\frac{\nabla _{\alpha}f'\nabla _{\beta}f'}{f'^{2}}+h_{\mu \nu}h^{\alpha \beta}\frac{2\nabla _{\alpha}\nabla _{\beta}f'}{3f'}~.
\end{align}
The only remaining part corresponds to the electric part $\mathbf{E}_{\mu \nu}$ of the bulk Weyl tensor $C_{abcd}$. From the transformation property of the Weyl tensor it immediately follows: $\mathbf{E}_{\mu \nu}=E_{\mu \nu}$. Combining all these, in the Einstein frame the effective gravitational field equations on the brane takes the form,
\begin{align}
\mathbf{T}_{\mu \nu}^{\rm brane}&=\mathbf{G}_{\mu \nu}-\left\lbrace \mathbf{K}\mathbf{K}_{\mu \nu}-\mathbf{K}^{\alpha}_{\mu}\mathbf{K}_{\nu \alpha}-\frac{1}{2}\mathbf{h}_{\mu \nu}\left(\mathbf{K}^{2}-\mathbf{K}_{\mu \nu}\mathbf{K}^{\mu \nu}\right)\right\rbrace +\mathbf{E}_{\mu \nu}-\kappa _{4}^{2}\mathbf{T}_{\mu \nu}^{(\phi)}
\nonumber
\\
&=G_{\mu \nu}-\left\lbrace KK_{\mu \nu}-K^{\alpha}_{\mu}K_{\nu \alpha}-\frac{1}{2}h_{\mu \nu}\left(K^{2}-K_{\mu \nu}K^{\mu \nu}\right)\right\rbrace +E_{\mu \nu}
\nonumber
\\
&-\frac{2}{3}\frac{\nabla _{\mu}\nabla _{\nu}f'}{f'}+h_{\mu \nu}\frac{f'R-f}{4f'}+h_{\mu \nu}h^{\alpha \beta}\frac{2\nabla _{\alpha}\nabla _{\beta}f'}{3f'}~,
\end{align}
which is precisely the effective gravitational field equations in the Jordan frame with the identification $T^{\rm brane}_{\mu \nu}=(1/f')\mathbf{T}_{\mu \nu}^{\rm brane}$. Hence the equivalence works at the level of effective field equations as well. However, the practical implementation of the above result again requires inversion of complicated functional forms and hence invites approximations.
\section{A Comparison with reconstruction methods in f(R) gravity}

In the above two sections we have shown the equivalence of gravitational field equations both in the bulk and in the brane respectively. In this section we will present a comparison of our method with an existing well known method in $f(R)$ gravity, the \emph{reconstruction method}. As already emphasized, due to presence of higher derivatives in the field equations for $f(R)$ gravity, obtaining a straightforward solution in a general case is very difficult. Even for systems with large number of symmetries, e.g., in cosmology which has a single unknown function $a(t)$, solving field equations \emph{directly} in the Jordan frame is very complicated. This lends its way to reconstruction method which we will briefly summarize \cite{Nojiri:2009kx,Nojiri:2006be,Nojiri:2009xh,Carloni:2010ph}.

In the reconstruction method one assumes that the expansion history of the universe is known exactly and by inverting the field equations one can determine what class of $f(R)$ theories can give rise to the observed universe. For example, power law solutions for the scale factor singles out $R^{n}$ to be the gravitational action. Since the scale factor $a(t)$ is known, the Ricci scalar is also known as $R(t)$. This can be inverted to get $t=g(R)$ and hence the Hubble parameter is known to be a function of Ricci scalar. This when used in the field equations, leads to a differential equation for $f(R)$, which can be solved to know the gravity model \cite{Nojiri:2009kx,Nojiri:2006be,Nojiri:2009xh}. There have been other variants of this model, e.g, assuming every physical quantity to be function of scale factor or function of Hubble parameter, which ultimately leads to a differential equation for the gravity model \cite{Carloni:2010ph}. The essential ingredients remain the same but one particular case may be 
convenient in comparison to the other in a particular situation. Let us now explicitly point out the advantages 
and disadvantages of reconstruction scheme as well as our approach.  
\begin{itemize}

\item  An important limitation of the reconstruction method is, only very simple cosmic histories, e.g. simple power law behaviours can be connected to $f(R)$ theory in an exact way. Our method, has similar disadvantages. Even though the field equations can be exactly solved in the Einstein frame for a few cases of interest, the inversion of the potential to $f(R)$ theory can be performed only in simple situations. 

\item The reconstruction method is adapted to cosmological spacetimes only, since the cosmic history is known through experiments. However the situation we are interested in corresponds to, higher dimensional physics in presence of higher curvature gravity, possible behaviour of the warp factor and the brane separation. Since there is no experimental backdrop for extra dimensions it is not possible to come up with a physical ansatz. Thus one needs to solve the field equations at face value, which can be efficiently done using our method as we have illustrated in the next sections. 

\item The utility of reconstruction scheme lies in its quick and straightforward analysis. Given a phenomenological scale factor $a(t)$ one needs to solve a single differential equation to get $f(R)$, given the inversion $t=t(R)$. While in our method one first need to solve the gravity plus scalar field system to get the solution in Einstein frame, which itself is a formidable task. Then one needs to invert the potential $V(\phi)$ to get $f(R)$ and finally the conformal transformation will yield the solution in Jordan frame. 

\end{itemize}
Thus both the methods have their own advantages and disadvantages. The reconstruction method is very simple, applicable even in presence of singularity and useful in cosmological context, while not so useful when applied to other scenarios e.g. extra dimensions. On the other hand, the method introduced in this work even though requires regular spacetime region, involves more steps to arrive at the solution, is very robust. It will work for any regular spacetime region, from cosmological scenarios to extra dimensions as explained in later examples. 
\section{Einstein to Jordan frame in the bulk: Explicit examples}\label{Sec_03}

We will now illustrate through simple examples how one might obtain solutions to bulk field equations in $f(R)$ gravity, which involve higher derivative terms, by exploiting the equivalence with scalar tensor theory depicted in the previous sections. As emphasized before, due to occurrence of higher derivative terms it is difficult to solve for the bulk equations of $f(R)$ gravity. On the other hand, solving a set of coupled equations of gravity plus scalar field system is much simpler. Hence through the equivalence shown earlier, if we can obtain a solution for the bulk metric $g_{ab}$ in the Einstein frame, corresponding solution in the Jordan frame will differ only by a conformal factor.

Before we jump into detailed calculations it is worthwhile to sketch the flowchart we are going to follow --- (a) We will start with the bulk action in the Einstein frame. (b) For some suitable potential, we will find out the metric describing bulk spacetime, by solving the bulk field equations. (c) We will match the potential in the Einstein frame with the corresponding $f(R)$ theory in the Jordan frame and finally (d) The conformal transformation will yield the corresponding bulk metric in the Jordan frame. In the examples to follow we will explicitly illustrate all the four steps mentioned above.

\paragraph*{Bulk field equations in Einstein frame} We start by solving the field equations of gravity and the scalar field in the Einstein frame. We assume that the branes are flat, viz., $\eta _{\mu \nu}$ is the spacetime metric on the brane. Further, the two branes are assumed to be separated by the stabilized value of the radion field $r_{c}$ \cite{Goldberger:1999uk,Chakraborty:2013ipa} such that the metric ansatz turns out to be (this ansatz is useful, particularly in the context of gauge hierarchy problem)
\begin{align}
d\mathbf{s}^{2}=e^{-2A(y)}\eta _{\mu \nu}dx^{\mu}dx^{\nu}+r_{c}^{2}dy^{2}~,
\end{align}
where $A(y)$ is the warp factor and is dependent on the extra spacetime coordinate alone. From the above metric ansatz the non-zero components of the Ricci tensor are immediate,
\begin{align}
\mathbf{R}_{\mu \nu}=\frac{e^{-2A}}{r_{c}^{2}}\left(A''-4A'^{2}\right)\eta _{\mu \nu};\qquad \mathbf{R}_{yy}=4A''-4A'^{2}~.
\end{align}
Note that we are following the previously mentioned convention: all the metric dependent quantities in the Einstein frame are boldfaced. From the components of the Ricci tensor, straightforward computation of the Ricci scalar leads to,
\begin{align}\label{Ricci}
\mathbf{R}=\frac{1}{r_{c}^{2}}\left(8A''-20A'^{2}\right)~.
\end{align}
Given the Ricci tensor and the Ricci scalar one can further compute the non-zero components of the Einstein tensor as,
\begin{align}
\mathbf{G}_{\mu \nu}=3\frac{e^{-2A}}{r_{c}^{2}}\eta _{\mu \nu}\left(-A''+2A'^{2}\right);\qquad \mathbf{G}_{yy}=6A'^{2}~,
\end{align}
such that the bulk gravitational field equations (see \ref{E_bulk}) in the Einstein frame become,
\begin{align}\label{bulk_E}
\mathbf{G}_{ab}=-\Lambda \mathbf{g}_{ab}+\kappa _{5}^{2}\left[\partial _{a}\phi \partial _{b}\phi -\mathbf{g}_{ab}\left\lbrace \frac{1}{2}\mathbf{g}^{cd}\partial _{c}\phi \partial _{d}\phi +V(\phi)\right\rbrace \right]~,
\end{align}
where, $\Lambda$ is the bulk cosmological constant. From \ref{Ricci} it is clear that bulk curvature depends only on the extra coordinate $y$, since $A$ depends on $y$ only. Thus logical consistency of the field equations demand that $\phi$ should also depend only on the extra dimensional coordinate $y$. In which case \ref{bulk_E} reduce to the following three coupled equations for gravity and scalar field as,
\begin{align}
-3A''+6A'^{2}&=-\Lambda r_{c}^{2}-\kappa _{5}^{2}\left[\frac{1}{2}\phi '^{2}+r_{c}^{2}V(\phi)\right]~,
\\
A'^{2}&=-\frac{\Lambda}{6}r_{c}^{2}+\frac{\kappa _{5}^{2}}{12}\phi '^{2}-\frac{r_{c}^{2}\kappa _{5}^{2}}{6}V(\phi)~,
\\
\phi ''-4A'\phi '&=r_{c}^{2}\frac{\partial V}{\partial \phi}~,
\end{align}
where `prime' denotes derivative with respect to $y$. Eliminating $A'$ from the first two equations we obtain, 
\begin{align}\label{Aprime}
A''=\frac{\kappa _{5}^{2}}{3}\phi '^{2}~.
\end{align}
In general the solution to the above coupled equations can be obtained by introducing a super-potential $W(\phi)$ which satisfies the following differential equation
\begin{align}
\frac{1}{9}W^{2}=-\frac{\Lambda}{6}r_{c}^{2}+\frac{1}{12\kappa _{5}^{2}}\left(\frac{\partial W}{\partial \phi}\right)^{2}-\frac{r_{c}^{2}\kappa _{5}^{2}}{6}V(\phi)~.
\end{align}
The above differential equation for $W(\phi)$ can be inverted and the potential $V(\phi)$ gets determined in terms of $W(\phi)$ as,
\begin{align}
V(\phi)=-\frac{\Lambda}{\kappa _{5}^{2}}+\frac{1}{2\kappa _{5}^{4}r_{c}^{2}}\left(\frac{\partial W}{\partial \phi}\right)^{2}-\frac{2}{3\kappa _{5}^{2}r_{c}^{2}}W^{2}~.
\end{align}
In such a scenario the coupled equations become separable. One thus obtains separate differential equations for the metric function $A$ and the scalar field $\phi$ in the following form
\begin{align}\label{field_eq}
A'=\frac{1}{3}W;\qquad \phi '=\frac{1}{\kappa _{5}^{2}}\frac{\partial W}{\partial \phi}~.
\end{align}
On the other hand, if one postulates the separability of the coupled field equations, then also the expression of the potential $V(\phi)$ in terms of the super-potential $W(\phi)$ follows. This choice for $A$ and $\phi$ also satisfies the field equation for $\phi$ as well, as one can easily check. The only remaining one corresponds to, \ref{Aprime}, i.e., the $A''$ equation. 

\paragraph*{Solutions in Einstein Frame} We have set the stage, it is now time to act. Having obtained the field equations in a sensible form, let us now solve for the bulk metric. In order to satisfy \ref{Aprime} one requires two possible choices for $W(\phi)$ --- (i) $W(\phi)=c\phi$, (ii) $W(\phi)=a-b\phi ^{2}$ for arbitrary choices of $a$, $b$ and $c$. Solving the field equations in both these cases separately leads to,
\begin{itemize}

\item  The super-potential is linear in $\phi$, i.e., $W(\phi)=c\phi$. Then from \ref{field_eq} one obtains, $A'=(b/3)\phi$, such that $A''=(c/3)\phi '$, while $\phi '=c/\kappa _{5}^{2}$. This set identically satisfies \ref{Aprime}. The corresponding potential $V(\phi)$ turns out to be, 
\begin{align}\label{V1}
V(\phi)=-\frac{\Lambda}{\kappa _{5}^{2}}+\frac{c^{2}}{2\kappa _{5}^{4}r_{c}^{2}}-\frac{2c^{2}}{3\kappa _{5}^{2}r_{c}^{2}}\phi ^{2}~,
\end{align}
with the following solution for $A(y)$ and $\phi (y)$ as,
\begin{align}
\phi (y)&=\phi _{0}+\frac{c}{\kappa _{5}^{2}}y~,
\\
A(y)&=A_{0}+\frac{c\phi _{0}}{3}y+\frac{c^{2}}{6\kappa _{5}^{2}}y^{2}~.
\end{align}

\item  The super-potential is quadratic in $\phi$, i.e., $W(\phi)=a-b\phi ^{2}$. From \ref{field_eq} we get, $A'=(1/3)(a-b\phi ^{2})$, hence this yields $A''=-(2b/3)\phi \phi '$, with $\phi '=-(2b/\kappa _{5}^{2})\phi$. These expressions can be easily manipulated to show that \ref{Aprime} is indeed satisfied. Then the potential becomes,
\begin{align}\label{V2}
V(\phi)=\left(-\frac{\Lambda}{\kappa _{5}^{2}}-\frac{2a^{2}}{3r_{c}^{2}\kappa _{5}^{2}}\right)+\left(\frac{b^{2}}{2r_{c}^{2}\kappa _{5}^{4}}+\frac{4ab}{3r_{c}^{2}\kappa _{5}^{2}}\right)\phi ^{2}-\frac{2b^{2}}{3r_{c}^{2}\kappa _{5}^{2}}\phi ^{4}~,
\end{align}
with the following solutions for $A(y)$ and $\phi (y)$,
\begin{align}
\phi (y)&=\phi _{0}\exp \left(-\frac{b}{\kappa _{5}^{2}}y\right)~,
\\
A(y)&=A_{0}+\sqrt{-\frac{\Lambda r_{c}^{2}}{6}}y+\frac{\kappa _{5}^{2}}{6}\phi _{0}^{2}\exp \left(-2\frac{b}{\kappa _{5}^{2}}y\right)~.
\end{align}
\end{itemize}
Thus we have exactly solved the bulk gravitational field equations in the Einstein frame for two choices of the scalar field potential. One of them is quadratic, i.e., $V(\phi)=a+b\phi ^{2}$ while the other corresponds to quartic potential, $V(\phi)=a+b\phi ^{2}+c\phi ^{4}$. Now we need to execute the last two steps in our flowchart, namely, (a) First one should identify a $f(R)$ model, which gives rise to the potentials obtained above and (b) Secondly, one needs to find the conformal factor relating the two frames and hence the metric can be obtained in the Jordan frame. 

\paragraph*{Connecting Jordan and Einstein frame} So far we have been working solely in the Einstein frame. In order to obtain the respective solution in the Jordan frame, we have to connect the scalar field to a $f(R)$ model, which can be done through the relations
\begin{align}\label{Neweq}
\kappa _{5}\phi =\frac{2}{\sqrt{3}}\ln f'(R);\qquad \kappa _{5}^{2}V=\frac{f'R-f}{2f'^{5/3}}~.
\end{align}
These relations follow from the original connection between Einstein and Jordan frame discussed in \ref{conditions_01} and \ref{conditions_02} respectively. In principle one should start with the quadratic and quartic potentials obtained previously and hence obtain the corresponding $f(R)$ theory in the Jordan frame by using \ref{Neweq}. However we will take the opposite route, i.e., we will start with some $f(R)$ model and arrive at the respective potentials in the Einstein frame using \ref{Neweq} and map it to those obtained earlier.
\begin{itemize}
 
\item The simplest model is always the best to start with. For $f(R)$ gravity this corresponds to a situation, where the Einstein-Hilbert term, receives a quadratic correction \footnote{The quadratic correction is well known in the literature, see for example \cite{Liu:2011wi,Barrow:2006xb,Barrow:2009gx,Barrow:1988xi,Maeda:1987xf}.}, i.e., $f(R)=R+\alpha R^{2}$. For this particular model of $f(R)$ gravity, the scalar field and the potential can be written in terms of the Ricci scalar in the Jordan frame as (using \ref{Neweq}),
\begin{align}
\kappa _{5}\phi &=\frac{2}{\sqrt{3}}\ln \left(1+2\alpha R\right);\qquad R=\frac{1}{2\alpha}\left[\exp\left({\frac{\sqrt{3}}{2}\kappa _{5}\phi}\right)-1\right]~,
\\
V(\phi)&=\frac{1}{8\alpha \kappa _{5}^{2}}\left[\exp \left(\frac{\kappa _{5}\phi}{2\sqrt{3}}\right)-2\exp \left(-\frac{\kappa _{5}\phi}{\sqrt{3}} \right)+\exp \left(-\frac{5}{2\sqrt{3}}\kappa _{5}\phi \right)\right]~.
\end{align}
The minima of the potential corresponds to $\partial V/\partial \phi =0$, with $\partial ^{2}V/\partial \phi ^{2}>0$. Both these conditions can be satisfied provided $e^{\kappa _{5}\phi}=1$, or $\phi =0$. Finally, expanding around the minima one obtains the following form for the potential,
\begin{align}\label{vphi_01}
V(\phi)&=\frac{1}{8\alpha \kappa _{5}^{2}}\left[\left(1+\frac{\kappa _{5}\phi}{2\sqrt{3}}+\frac{1}{2}\frac{\kappa _{5}^{2}\phi ^{2}}{12} \right)-2\left(1-\frac{\kappa _{5}\phi}{\sqrt{3}}+\frac{1}{6}\kappa _{5}^{2}\phi ^{2}\right)+\left(1-\frac{5}{2\sqrt{3}}\kappa _{5}\phi +\frac{1}{2}\frac{25\kappa _{5}^{2}\phi ^{2}}{12}\right)\right]
\nonumber
\\
&=\frac{3}{32\alpha}\phi ^{2}~.
\end{align}
Thus we have a quadratic potential for $\phi$, which originate from $R+\alpha R^{2}$ gravity. Matching the potential to that derived in the Einstein frame, given by \ref{V1}, we obtain the following relation: $\Lambda =-9/(128\alpha)$.

\item Let us now consider a more general $f(R)$ gravity model for which $f(R)=R+\alpha R^{2}+\beta R^{4}$, where $\alpha$ and $\beta$ are dimension full constant coefficients, with values such that the model becomes ghost free. Then from \ref{Neweq} we obtain, $R=(\sqrt{3}\kappa _{5}\phi)/(4\alpha)$, such that the potential turns out to be, 
\begin{align}
V(\phi)=\frac{\alpha R^{2}+3\beta R^{4}}{2\kappa _{5}^{2}}=\frac{3}{32\alpha}\phi ^{2}+\frac{1}{2}\frac{3^{3}\beta \kappa _{5}^{2}}{4^{4}\alpha ^{4}}\phi ^{4}~,
\end{align}
where we have assumed $\alpha \gg \beta \gg \alpha ^{2}$, consistent with the ghost free criteria for this $f(R)$ model. Comparing this with the potential obtained by solving bulk field equations in the Einstein frame, presented in \ref{V2} we immediately obtain, 
\begin{align}
a=\sqrt{-\frac{3r_{c}^{2}\Lambda}{2}};\qquad b=-\frac{4a\kappa _{5}^{2}}{3}\pm \sqrt{\frac{16a^{2}\kappa _{5}^{4}}{9}+\frac{3r_{c}^{2}\kappa _{5}^{4}}{16\alpha}};\qquad \beta =-\frac{4^{5}b^{2}\alpha^{4}}{3^{4}r_{c}^{2}\kappa _{5}^{4}}~.
\end{align}
\end{itemize}
This completes the connection between Einstein and Jordan frames. The potentials obtained in the Einstein frame get mapped to respective $f(R)$ theories. 

\paragraph*{Solutions in Jordan Frame} We have now reached the final step of our flowchart, viz., solution to the bulk field equations in the Jordan frame. For this purpose we can use the connection with Einstein frame derived earlier. 
\begin{itemize}

\item For $f(R)=R+\alpha R^{2}$, the corresponding scalar field potential in the Einstein frame is quadratic with the mapping being given by \ref{vphi_01}. From which the conformal factor turns out to be, $\Omega=(1+2\alpha R)^{1/3}=[1+(\sqrt{3}\kappa _{5}\phi/2)]^{1/3}$. Hence the bulk solution in the Jordan frame corresponds to,
\begin{align}\label{fr1}
ds^{2}&=\left[1+\frac{\sqrt{3}\kappa _{5}\phi(y)}{2}\right]^{-2/3}\left\lbrace e^{-2A(y)}\eta _{\mu \nu}dx^{\mu}dx^{\nu}+r_{c}^{2}dy^{2}  \right\rbrace ~,
\\
\phi (y)&=\phi _{0}+\frac{c}{\kappa _{5}^{2}}y;\qquad A(y)=A_{0}+\frac{c\phi _{0}}{3}y+\frac{c^{2}}{6\kappa _{5}^{2}}y^{2}~,
\nonumber
\end{align}
where $c$ and $\phi_{0}$ are arbitrary constants of integration. Thus for the quadratic $f(R)$ model under consideration, one can map it to the Einstein frame and obtain the respective potential. For this particular case, the field equations in the Einstein frame becomes exactly solvable and hence by conformal transformation one can obtain the corresponding solution in the Jordan frame. Further, from \ref{fr1} it turns out that the warp factor is governed by the factors $c\phi _{0}$ and $c/\kappa _{5}$. Hence in order to have proper suppression of the Planck scale on the visible brane one must have the conditions $c<\kappa _{5}$ and $c\phi _{0}\sim 36$. Hence one arrives at, $\phi_{0}>\kappa _{5}^{-1}$. Further, in this model the radion field varies with extra dimension $y$ as $(a+by)^{2/3}$, where $a$ and $b$ depends on $c,\kappa _{5}$ and $\phi _{0}$.

\item For the other model, i.e., $f(R)=R+\alpha R^{2}+\beta R^{4}$, the scalar field potential in the Einstein frame is quartic. From which one can relate the parameters $\alpha,\beta$ with the respective ones in the Einstein frame. Following the same strategy as above, the conformal factor turns out to yield, $\Omega =(1+2\alpha R+4\beta R^{3})^{1/3}=[1+(\sqrt{3}\kappa _{5}\phi/2)+(3\sqrt{3}\beta \kappa _{5}^{3}\phi ^{3}/16\alpha ^{3})]^{1/3}$. Using which the solution in the Jordan frame becomes,
\begin{align}\label{fr2}
ds^{2}&=\left[1+\frac{\sqrt{3}\kappa _{5}\phi(y)}{2}+\frac{3\sqrt{3}\beta \kappa _{5}^{3}\phi(y)^{3}}{16\alpha ^{3}}\right]^{-2/3}
\left\lbrace e^{-2A(y)}\eta _{\mu \nu}dx^{\mu}dx^{\nu}+r_{c}^{2}dy^{2} \right\rbrace ~,
\\
\phi (y)&=\phi _{0}\exp \left(-\frac{b}{\kappa _{5}^{2}}y\right);\qquad A(y)=A_{0}+\sqrt{-\frac{\Lambda r_{c}^{2}}{6}}y+\frac{\kappa _{5}^{2}}{6}\phi _{0}^{2}\exp \left(-2\frac{b}{\kappa _{5}^{2}}y\right)~,
\nonumber
\end{align}
This demonstrates another $f(R)$ theory for which the corresponding potential in the Einstein frame leads to an exact solution. Using which and the mapping between Einstein and Jordan frame one obtains the respective solution in the Jordan frame. However in contrast with the previous situation, in this case the warp factor behaves exactly like the Randall-Sundrum scenario, since all the corrections are exponentially suppressed (see \ref{fr2}).  While the radion field is almost constant due to identical exponential suppression. Hence the $f(R)$ model with quartic correction is more favored in the extra dimensional physics than the earlier one. 

\end{itemize}

\paragraph*{Aside: Comment on domain of applicability} After illustrating two examples on how to obtain solutions to higher order field equations, by using the Einstein frame judiciously, we would like to comment on the domain of applicability of this approach.

We should emphasize that we are working in a mesoscopic energy scale, i.e., the energy scale is larger compared to general relativity, such that effect of higher order terms, e.g, $\alpha R^{2}$ cannot be ignored. On the other hand, the energy scale is much smaller compared to the Planck scale so that the additional contributions are still sub-dominant, i.e., $\alpha R<1$. This is important, in particular when one obtains the scalar field in the Einstein frame in terms of the curvature. In order to obtain closed form expression one has to expand the potential near its minimum, this in turn requires one to neglect higher order curvature corrections, e.g., one might neglect $\alpha ^{2}R^{2}$ in comparison with $\alpha R$. In a nutshell, we are working in a high curvature regime such that effect of $f(R)$ gravity can be felt, but not high enough so that the Einstein-Hilbert action becomes sub-dominant.

Another point that requires clarification are the approximations involved in general scenarios. The conformal transformation and hence the conversion of a potential to a corresponding $f(R)$ model is not at all straightforward. In most of the cases the relations turn out to be non-invertible, and one need to resort to approximations. As explained earlier, on physical grounds, one can assume that the higher order terms are sub-leading and hence one can keep only linear order terms. While dealing with complicated potentials most often one needs to resort to these approximations, justified by physical intuitions. However at the Planck scale these approximations brake down, since the assumption that higher orders terms are sub-leading cannot be trusted.

Having discussed two possible scenarios in the context of bulk physics let us now consider brane dynamics. In particular, we will be interested in one spherically symmetric and one cosmological applications.
\section{Jordan to Einstein frame in the brane: Explicit examples}\label{Sec_04}

Both the examples depicted above are related to bulk spacetimes. To complete the discussion we will also derive the metric in the Einstein frame starting from the Jordan frame, but in the brane spacetime. This is to explicitly demonstrate that the technique works both ways --- whenever it is convenient to solve in the Einstein frame, we can solve it and transform back to the Jordan frame, while if the solution is simpler in Jordan frame it can give insight into what happens in scalar coupled gravity, viz., the Einstein frame. 
\begin{itemize}
 
\item In this example, we will start with a particular $f(R)$ model on the brane, solve the effective field equations and obtain a cosmological solution. Then using conformal transformation the corresponding solution in the Einstein frame can be obtained. 
\paragraph*{Solution in the Jordan frame} Let us start with the $f(R)$ model given by $f(R)=f_{0}(R-R_{0})^{\alpha}$, where $f_{0}$ and $R_{0}$ are constants and $\alpha \neq 1$. The corresponding solution for the scale factor on the brane can be obtained by solving the effective field equations derived in \cite{Chakraborty:2014xla,Haghani:2012zq}. This leads to power law solution $a(t)\sim t^{n}$, where $n$ is related to $\alpha$ and the matter fields present on the brane. 
\paragraph*{Converting back to Einstein frame} We need to convert it back to Einstein frame and hence obtain the corresponding solution in the scalar coupled gravity. For this choice for $f(R)$, we obtain, the scalar field to be,
\begin{align}
\kappa _{5}\phi =\frac{2}{\sqrt{3}}\ln \left(f_{0}\alpha \right)+\frac{2}{\sqrt{3}}\left(\alpha -1\right)\ln \left(R-R_{0}\right)~,
\end{align}
which in turn can be inverted, leading to, $R-R_{0}=\exp[a\kappa _{5}(\phi -\phi_{0})]$, where, $a=\sqrt{3}/(2(\alpha -1))$ and $\kappa _{5}\phi _{0}=(2/\sqrt{3})\ln (f_{0}\alpha)$. Then the potential can be determined readily, using \ref{conditions_02}, leading to,
\begin{align}
V(\phi)=\frac{\alpha -1}{2\kappa _{5}^{2}f_{0}^{2/3}\alpha ^{5/3}}\exp\left[\kappa _{5}a\left(\frac{5}{3}-\frac{2\alpha}{3}\right)\left(\phi -\phi _{0}\right) \right]~.
\end{align}
Hence the power law behavior of the $f(R)$ theory transforms back to exponential potential. 
\paragraph*{Solution in Einstein frame} The corresponding cosmological solution in the Einstein frame could be obtained by transforming the Jordan frame solution using the conformal factor. In this particular class of $f(R)$ model, the conformal factor becomes, $\Omega =(f_{0}\alpha)^{1/3}(R-R_{0})^{(\alpha -1)/3}\sim t^{-2(\alpha -1)/3}$. Thus the corresponding solution in the Einstein frame is again cosmological with a new scale factor: $\hat{a}(t)\sim t^{[3n-4(\alpha -1)]/3}$. Hence the cosmological solution in the Einstein frame with exponential potential is still power law.

\item So far we have been dealing with power law $f(R)$ theories. However to show the applicability of our method to more general scenarios, we will consider the following $f(R)$ model: $R-(\alpha/R)$ on the brane. This $f(R)$ model in four dimensional spacetime has been discussed in detail in \cite{Nojiri:2003ft}, however no such solution in the context of effective field equations exist, which by itself would be an interesting future work. However in this work we will contend ourselves in providing basic ingredients regarding this model. Further given a solution in Jordan frame, use of conformal transformation will lead to the corresponding solution in the Einstein frame. 
\paragraph*{Solution in the Jordan frame} Let us start with the above mentioned $f(R)$ model. The corresponding solution for the scale factor on the brane can be obtained by solving the effective field equations and can be taken to be $a(t)\sim t^{n}$, where $n$ should be related to $\alpha$ and the matter fields present on the brane. 
\paragraph*{Converting back to Einstein frame} We need to convert it back to Einstein frame and hence obtain the corresponding solution in the scalar coupled gravity. For this choice for $f(R)$, we obtain, the scalar field to be,
\begin{align}
\kappa _{5}\phi =\frac{2}{\sqrt{3}}\ln \left(1+\frac{\alpha}{R^{2}}\right)~,
\end{align}
which in turn can be inverted, leading to, $R=\sqrt{\alpha}(\exp[(2/\sqrt{3})\kappa _{5}\phi]-1)^{-1/2}$. Then the potential can be determined readily, using \ref{conditions_02}, leading to,
\begin{align}
V(\phi)=\sqrt{\alpha}\exp\left(-\frac{5}{2\sqrt{3}}\kappa _{5}\phi\right)\sqrt{\exp[(2/\sqrt{3})\kappa _{5}\phi]-1}~.
\end{align}
Expanding for small $\phi$, we obtain, $V(\phi)=\sqrt{\alpha \kappa _{5}\sqrt{3}\phi/2}~(1-(5\kappa _{5}\phi/2\sqrt{3}))$.
Hence negative power law behavior of the $f(R)$ theory transforms back to potential with $\sqrt{\phi}$ as the leading order contribution. 
\paragraph*{Solution in Einstein frame} The corresponding cosmological solution in the Einstein frame could be obtained by transforming the Jordan frame solution using the conformal factor. In this particular class of $f(R)$ model, the conformal factor becomes, $\Omega =[1+(\alpha/R^{2})]^{1/3}\sim [1+\alpha t^{4}]^{1/3}$, since $R\sim t^{-2}$. Thus for late times, $\Omega \sim t^{4/3}$. Thus the corresponding late time solution in the Einstein frame is again cosmological with a new scale factor: $\hat{a}(t)\sim t^{n+(8/3)}$, again a power law behavior. Hence the cosmological solution in the Einstein frame with $\sqrt{\phi}$ potential is still a power law.

\item Another explicit spherically symmetric solution on the brane in the Jordan frame has been constructed in \cite{Chakraborty:2014xla} by decomposing the Electric part of the Weyl tensor into dark radiation term $U(r)$ and dark pressure term $P(r)$.
\paragraph*{Solution in Jordan frame} The dark radiation $U(r)$ and dark pressure $P(r)$ acts as auxiliary source to the effective gravitational field equations \cite{Maartens:2001jx}. A possible solution can be obtained when an ``equation of state'' between $U(r)$ and $P(r)$ is specified. For the particular choice $2U(r)+P(r)=0$, we immediately obtain, the corresponding 
spherically symmetric solution \cite{Chakraborty:2014xla},
\begin{align}\label{frmetric}
ds^{2}=-f(r)dt^{2}+\frac{dr^{2}}{f(r)}+r^{2}d\Omega ^{2};\qquad f(r)=1-\frac{2GM+Q_{0}}{r}-\frac{3\bar{\kappa}P_{0}}{2r^{2}}+\frac{F(R)-\Lambda _{4}}{3}r^{2}~.
\end{align}
Here $Q_{0}$ and $P_{0}$ corresponds to constants of integration, $\bar{\kappa}$ captures the effect of bulk spacetime, i.e., depends on bulk gravitational constant and $F(R)$, evaluated at the brane location can be constructed from the original $f(R)$ theory by taking appropriate derivative. Assuming that the bulk scalar depends only on the bulk coordinates and for a $f(R)$ theory of the form $f(R)=R+\alpha R^{2}+\beta R^{4}$, the leading order behavior of $F(R)$ is like an effective four dimensional cosmological constant $\Lambda _{4}$.
\paragraph*{Converting back to Einstein frame} The corresponding scalar-tensor solution can be obtained by transforming the metric in \ref{frmetric} using the appropriate conformal factor: $\Omega =(1+2\alpha R+4\beta R^{3})^{1/3}=[1+(3\kappa _{5}\phi/2)+(27\beta \kappa _{5}^{3}\phi ^{3}/64\alpha ^{3})]^{1/3}$. Since the scalar field depends on the extra coordinate only, the conformal factor evaluated on the location of the brane is just a constant.  
\paragraph*{Solution in Einstein frame} Since the corresponding conformal factor is just a constant it will scale the metric, which can be absorbed by rescaling of time and radial coordinate by the conformal factor. Hence the solution in the Einstein frame would remain the same. 

\end{itemize}
Thus even in the context of brane models solving effective gravitational field equations in one frame and obtaining the solution in other often requires approximations. One has to keep in mind that we are working in the mesoscopic scale, where neither the higher curvature terms are dominant nor are they negligible. This allows one to invert various relations connecting the Einstein frame scalar with Jordan frame curvature, a key aspect while converting solutions in one frame to another.
\section{Conclusions}

A technique for solving field equations of higher curvature gravity theories have been proposed. The technique essentially hinges on the mathematical equivalence of higher curvature gravity theory, e.g., $f(R)$ theories of gravity with scalar-tensor representation. Earlier this equivalence was known only at the level of action principle. It was not clear a priori whether the field equations derived from either the $f(R)$ representation or the scalar-tensor representation would also be equivalent. In this work starting from a five dimensional theory we have explicitly demonstrated --- (a) the bulk gravitational field equations derived from Jordan and Einstein frame are equivalent and (b) the effective field equations on the brane in these two approaches are also equivalent. Using this equivalence we have argued, if one can solve for the field equations in one frame, the solution in the other frame can be easily obtained. Even though for simple models one can perform the above operation exactly, often it 
requires suitable approximations. The approximations essentially requires one to work in mesoscopic energy scales, viz., higher than weak scale but less than Planck scale. For practical application of the technique, we have illustrated it in two related situations:
\begin{itemize}

\item The bulk field equation in the Einstein frame have been solved in the context of warped geometry models for two choices of the potential --- quadratic and quartic. Following expectation, the warp factor behaves differently in these two scenarios, but leads to desired exponential warping. These potential through conformal transformations are related to two $f(R)$ models --- (a) $R +\alpha R^{2}$ and (b) $R+\alpha R^{2}+\beta R^{4}$ respectively. From the solution in the Einstein frame we have obtained the solution in $f(R)$ representation as well, having a different warp factor behavior and extra dimension dependent radion field.

\item Secondly, using the known solutions to effective field equations in the $f(R)$ representations we have obtained the corresponding solutions in the scalar-tensor representation. In the cosmological context, the scale factor still exhibit power law behavior, but with a different power. While the spherically symmetric solution results in mere rescaling of the coordinates.
 
\end{itemize}
In the two examples depicted in this work we have explored practical illustration of the technique for both the bulk and brane spacetimes. Further it turns out that even though in the Einstein frame the brane separation has been fixed at the stabilized value, in the Jordan frame it starts depending on the extra dimension. Interestingly, the warp factor in the two frames are different, leading to different suppression of the Planck scale on the visible brane. This might lead to potential observables, distinguishing the two frames in the context of recent LHC experiments.
\section*{Acknowledgements}

Research of S.C. is funded by a SPM fellowship from CSIR, Government of India. He also thanks IACS, India for warm hospitality; a part of this work was completed there during a visit. 
\appendix
\labelformat{section}{Appendix #1} 
\labelformat{subsection}{Appendix #1}
\section{Appendix: Conformal Transformation}\label{GBAPP_01}

In what follows conformal transformations will play a major role throughout our discussion. Thus before jumping into the main body of this work, let us briefly review some necessary ingredients of conformal transformation. 

Let us start by reviewing conformal transformation in which the metric $g_{ab}\rightarrow \mathbf{g}_{ab}$, such that,
\begin{align}
\mathbf{g}_{ab}=\Omega ^{2}g_{ab}
\end{align}
Under this transformation the curvature tensor and its various contractions in $D$-dimensional spacetime transform as,
\begin{subequations}
\begin{align}
\mathbf{R}^{a}_{~bcd}&=R^{a}_{~bcd}-2\left(\delta ^{a}_{[c}\delta ^{e}_{d]}\delta ^{f}_{b}-g_{b[c}\delta ^{e}_{d]}g^{af}\right)\frac{1}{\Omega}\nabla _{e}\nabla _{f}\Omega +2\left(2\delta ^{a}_{[c}\delta ^{e}_{d]}\delta ^{f}_{b}-2g_{b[c}\delta ^{e}_{d]}g^{af}+g_{b[c}\delta ^{a}_{d]}g^{ef}\right)\frac{1}{\Omega ^{2}}\nabla _{e}\Omega \nabla _{f}\Omega
\label{Rabcd_Transform}
\\
\mathbf{R}_{ab}&=R_{ab}-\left\lbrace \left(D-2\right)\delta ^{e}_{a}\delta ^{f}_{b}+g_{ab}g^{ef}\right\rbrace \frac{1}{\Omega}\nabla _{e}\nabla _{f}\Omega +\left\lbrace 2\left(D-2\right)\delta ^{e}_{a}\delta ^{f}_{b}-\left(D-3\right)g_{ab}g^{ef}\right\rbrace \frac{1}{\Omega ^{2}}\nabla _{e}\Omega \nabla _{f}\Omega
\label{Rab_Transform}
\\
\mathbf{R}&=\frac{R}{\Omega ^{2}}-2\left(D-1\right)g^{ef}\frac{1}{\Omega ^{3}}\nabla _{e}\nabla _{f}\Omega-\left(D-1\right)\left(D-4\right)g^{ef}\frac{1}{\Omega ^{4}}\nabla _{e}\Omega \nabla _{f}\Omega
\label{Ric_Transform}
\end{align}
\end{subequations}
while Weyl tensor remains invariant under conformal transformation, such that,
\begin{align}
\mathbf{C}_{abcd}=\Omega ^{2}C_{abcd}
\end{align}
Let us concentrate on bulk spacetime, which we assume to have dimension $D=5$. Then Ricci scalar in the frame $g_{ab}$ (known as Jordan frame) is related to the Ricci scalar in the frame $\mathbf{g}_{ab}$ (known as Einstein frame) as,
\begin{align}\label{R_Transform}
R&=\Omega ^{2}\mathbf{R}-8\Omega ^{3}\mathbf{g}^{ab}\nabla _{a}\nabla _{b}\left(\frac{1}{\Omega}\right)-4\Omega ^{4}\mathbf{g}^{ab}\nabla _{a}\left(\frac{1}{\Omega}\right)\nabla _{b}\left(\frac{1}{\Omega}\right)
\nonumber
\\
&=\Omega ^{2}\mathbf{R}-8\Omega ^{3}\mathbf{g}^{ab}\nabla _{a}\left(-\frac{\nabla _{b}\ln \Omega}{\Omega}\right)-4\mathbf{g}^{ab}\nabla _{a}\Omega \nabla _{b}\Omega
\nonumber
\\
&=\Omega ^{2}\mathbf{R}+8\Omega ^{2}\boldsymbol{\square}\ln \Omega -12\mathbf{g}^{ab}\nabla _{a}\Omega \nabla _{b}\Omega
\end{align}
These are the results used in \ref{Sec_01}. 

To get the effective field equations in $(D-1)$-dimensional spacetime, one needs to concentrate on the transformation property for the extrinsic curvature. We will start with the transformation of extrinsic curvatures first,
\begin{align}
\mathbf{K}_{ab}=\mathbf{h}^{c}_{a}\mathbf{h}^{d}_{b}\boldsymbol{\nabla}_{c}\mathbf{n}_{d}=\mathbf{h}^{c}_{a}\mathbf{h}^{d}_{b}
\left(\partial _{c}\mathbf{n}_{d}-\boldsymbol{\Gamma}^{m}_{cd}\mathbf{n}_{m}\right)
\end{align}
Let us now concentrate on the normal $n_{a}$, for which, in the Einstein frame we have,
\begin{align}
\mathbf{n}_{a}=\Omega n_{a};\qquad \mathbf{n}^{a}=\frac{1}{\Omega}n^{a}
\end{align}
It is clear that since $n_{a}n^{a}=1$, we have, $\mathbf{n}_{a}\mathbf{n}^{a}=1$ as well. This leads to the result $\mathbf{h}^{a}_{b}=h^{a}_{b}$. Further the connection, under conformal transformation transforms as,
\begin{align}
\mathbf{\Gamma}^{m}_{cd}=\Gamma ^{m}_{cd}+\delta ^{m}_{c}\partial _{d}\ln \Omega +\delta^{m}_{d}\partial _{c}\ln \Omega -g_{cd}g^{mn}\partial _{n}\ln \Omega
\end{align}
This immediately leads to,
\begin{align}\label{Kab_Transform}
\mathbf{K}_{ab}&=h^{c}_{a}h^{d}_{b}\left[\partial _{c}\left(\Omega n_{d}\right)-\Omega n_{m}\left(\Gamma ^{m}_{cd}+\delta ^{m}_{c}\partial _{d}\ln \Omega +\delta^{m}_{d}\partial _{c}\ln \Omega -g_{cd}g^{mn}\partial _{n}\ln \Omega\right)\right]
\nonumber
\\
&=\Omega K_{ab}+h_{ab}\left(n^{c}\partial _{c}\Omega \right)
\end{align}
Using the above transformation for the extrinsic curvature we can obtain easily the following transformations as well,
\begin{align}
\mathbf{K}^{a}_{b}&=\mathbf{g}^{ac}\mathbf{K}_{cb}=\frac{1}{\Omega}K^{a}_{b}+\frac{1}{\Omega ^{2}}h^{a}_{b}\left(n^{c}\partial _{c}\Omega \right)
\\
\mathbf{K}&=\frac{1}{\Omega}K+\frac{D-1}{\Omega ^{2}}\left(n^{c}\partial _{c}\Omega \right)
\\
\mathbf{K}^{ab}&=\frac{1}{\Omega ^{3}}K^{ab}+\frac{1}{\Omega ^{4}}h^{ab}\left(n^{c}\partial _{c}\Omega \right)
\end{align}
where we have assumed the bulk spacetime to have a dimension $D$. After providing the general result we will confine ourselves with $D=5$. These lead to the following relation 
\begin{align}
\mathbf{K}\mathbf{K}_{\mu \nu}&=\left[\frac{1}{\Omega}K+\frac{D-1}{\Omega ^{2}}\left(n^{c}\partial _{c}\Omega \right)\right]
\left[\Omega K_{\mu \nu}+h_{\mu \nu}\left(n^{c}\partial _{c}\Omega \right)\right]
\nonumber
\\
&=KK_{\mu \nu}+(D-1)K_{\mu \nu}\left(n^{c}\partial _{c}\ln \Omega \right)+Kh_{\mu \nu}\left(n^{c}\partial _{c}\ln \Omega \right)+(D-1)h_{\mu \nu}\left(n^{c}\partial _{c}\ln \Omega \right)^{2}
\\
\mathbf{K}^{\alpha}_{\mu}\mathbf{K}_{\nu \alpha}&=\left[\frac{1}{\Omega}K^{\alpha}_{\mu}+\frac{1}{\Omega ^{2}}h^{\alpha}_{\mu}\left(n^{c}\partial _{c}\Omega \right)\right]\left[\Omega K_{\nu \alpha}+h_{\nu \alpha}\left(n^{c}\partial _{c}\Omega \right)\right]
\nonumber
\\
&=K^{\alpha}_{\mu}K_{\nu \alpha}+2K_{\mu \nu}\left(n^{c}\partial _{c}\ln \Omega \right)+h_{\mu \nu}\left(n^{c}\partial _{c}\ln \Omega \right)^{2}
\\
\mathbf{K}^{2}&=\left[\frac{1}{\Omega}K+\frac{D-1}{\Omega ^{2}}\left(n^{c}\partial _{c}\Omega \right)\right]^{2}
\nonumber
\\
&=\frac{1}{\Omega ^{2}}\left[K^{2}+2(D-1)K\left(n^{c}\partial _{c}\ln \Omega \right)+(D-1)^{2}\left(n^{c}\partial _{c}\ln \Omega \right)^{2}\right]
\end{align}
\begin{align}
\mathbf{K}_{\mu \nu}\mathbf{K}^{\mu \nu}&=\left[\Omega K_{\mu \nu}+h_{\mu \nu}\left(n^{c}\partial _{c}\Omega \right)\right]
\left[ \frac{1}{\Omega ^{3}}K^{\mu \nu}+\frac{1}{\Omega ^{4}}h^{\mu \nu}\left(n^{c}\partial _{c}\Omega \right)\right]
\nonumber
\\
&=\frac{1}{\Omega ^{2}}\left[K_{\mu \nu}K^{\mu \nu}+2K\left(n^{c}\partial _{c}\ln \Omega \right)+(D-1)\left(n^{c}\partial _{c}\ln \Omega \right)^{2}\right]
\end{align}
Combining all these relations we finally arrive at,
\begin{align}\label{final_K}
\mathbf{K}\mathbf{K}_{\mu \nu}&-\mathbf{K}^{\alpha}_{\mu}\mathbf{K}_{\nu \alpha}-\frac{1}{2}\mathbf{h}_{\mu \nu}
\left(\mathbf{K}^{2}-\mathbf{K}_{\mu \nu}\mathbf{K}^{\mu \nu}\right)=KK_{\mu \nu}+(D-1)K_{\mu \nu}\left(n^{c}\partial _{c}\ln \Omega \right)+Kh_{\mu \nu}\left(n^{c}\partial _{c}\ln \Omega \right)
\nonumber
\\
&+(D-1)h_{\mu \nu}\left(n^{c}\partial _{c}\ln \Omega \right)^{2}-K^{\alpha}_{\mu}K_{\nu \alpha}-2K_{\mu \nu}\left(n^{c}\partial _{c}\ln \Omega \right)-h_{\mu \nu}\left(n^{c}\partial _{c}\ln \Omega \right)^{2}
\nonumber
\\
&-\frac{1}{2}h_{\mu \nu}\Big[K^{2}+2(D-1)K\left(n^{c}\partial _{c}\ln \Omega \right)+(D-1)^{2}\left(n^{c}\partial _{c}\ln \Omega \right)^{2}-K_{\mu \nu}K^{\mu \nu}
\nonumber
\\
&-2K\left(n^{c}\partial _{c}\ln \Omega \right)-(D-1)\left(n^{c}\partial _{c}\ln \Omega \right)^{2}\Big]
\nonumber
\\
&=KK_{\mu \nu}-K^{\alpha}_{\mu}K_{\nu \alpha}-\frac{1}{2}h_{\mu \nu}\left(K^{2}-K_{\mu \nu}K^{\mu \nu}\right)
+(D-3)\left(K_{\mu \nu}-Kh_{\mu \nu}\right)\left(n^{c}\partial _{c}\ln \Omega \right)
\nonumber
\\
&-\frac{(D-3)(D-2)}{2}h_{\mu \nu}\left(n^{c}\partial _{c}\ln \Omega \right)^{2}
\end{align}
In this the term linear in $n^{c}\nabla _{c}\ln \Omega$ has no effect, since this can be eliminated using the surface term of \ref{E_Action_02}, which is also linear in $n^{c}\nabla _{c}\ln \Omega$. These results are used in \ref{Sec_02} with $D=5$. 
\bibliography{Brane,Gravity_1_full,Gravity_2_partial}

\bibliographystyle{./utphys1}
\end{document}